\documentclass[12pt]{article}
\usepackage{amssymb,epsfig,setspace}
\begin{document}

\title{Non-radiative decay of 
a dipole emitter close to a metallic nanoparticle: 
Importance of higher-order multipole contributions
}

\author{
Alexander Moroz\thanks{http://www.wave-scattering.com}
\\
Wave-scattering.com
} 

\date{}

\maketitle

\begin{center}
{\large\sc abstract}
\end{center}
The contribution of higher-order multipoles to radiative and
non-radiative decay of a single dipole emitter close to a spherical
metallic nanoparticle is re-examined. Taking a Ag spherical
nanoparticle (AgNP) with the radius of $5$ nm as an example, 
a significant contribution (between $50\%$ and $101\%$ of the total
value) of higher-order multipoles to non-radiative rates is
found even at the emitter distance of 
$5$ nm from the AgNP surface. On the other hand,
the higher-order multipole contribution to radiative
rates is negligible. Consequently, a dipole-dipole approximation
can yield only an upper bound on the apparent quantum yield.
In contrast, the non-radiative rates calculated with 
the quasistatic Gersten and Nitzan method 
are found to be in much better
agreement with exact electrodynamic results. Finally, the size
corrected metal dielectric function is shown to decrease the
non-radiative rates near the dipolar surface plasmon resonance.

\vspace*{0.6cm}

PACS numbers:
33.50.-j, 33.50.Hv, 32.50.+d, 78.67.-n, 71.45.Gm, 73.22.Lp, 78.67.Bf \hfill

\vspace*{1.9cm}

\newpage

\section{Introduction}
Metallic surfaces are known to increase the local 
electromagnetic fields and modify 
both excitation and emission rates of proximate fluorophores, 
chromophores, and quantum dots 
\cite{Rup,DCL,Lak,Dulk,Lak5,AMap,DRK,AMcl,JSS,CGH,ABN,SRW}.
Especially promising is modifying fluorescence rates
by means of metallic nanoparticles (MNPs) for
addressing various issues of interest to biology \cite{DCL,Lak,Dulk,Lak5}. 
The MNPs can quench fluorescence as much as $100$ times better than 
other quenchers of fluorescence, such as
DABCYL, and open new perspectives in the use of hybrid materials 
as sensitive probes in fluorescence-based detection assays \cite{DCL}.
Molecular beacons comprising the MNP acceptors can detect
minute amounts of oligonucleotide target sequences in a pool of
random sequences and provide an
improved detection of a single mismatch in a competitive
hybridization assay for DNA mismatch detection \cite{DCL}.
The MNP acceptors can also be employed
for probing changes in distances for 
protein interactions on DNA using a molecular ruler approach 
called nanometal surface energy transfer \cite{JSS}.
The MNPs induced decrease in lifetime of fluorophores can result in an 
effective increase in photostability. All that carries
potential for the next generation superstructures in
clinical diagnostic, DNA sequencing, 
genomics, and biological detection. The latter is the subject
of a new field of radiative decay engineering that
aims at a predetermined modification of 
decay rates \cite{Lak,Lak5}.

In theory, fluorophores, chromophores, and quantum dots are 
modeled as dipole emitters
and the underlying problem is then that of a single 
dipole emitter in proximity to a MNP 
\cite{Rup,Dulk,AMap,DRK,AMcl,CGH,ABN,SRW}. 
In general, when the influence of a particle
on the dipole emission is studied, the {\em outgoing} 
electromagnetic fields of a dipole emitter are expanded into 
the electromagnetic multipole fields centered at a origin of and {\em incident} 
on the particle 
\cite{Rup,AMap,AMcl,Ch,Chew,KLG2}. This transformation 
induces formally an infinite number of higher order multipole 
electromagnetic fields that can be characterized
by the angular momentum numbers $l$ and $-l \leq m\leq l$ 
\cite{Rup,AMap,AMcl,Ch,Chew,KLG2}. 
A recent publication by Carminati et al \cite{CGH} suggested
an approximate analytical approach  to 
calculate the fluorescence decay rate and the
radiative and non-radiative rates. The rates were determined solely 
as a function of the particle polarizability $\alpha$.
For instance, the {\em fluorescence decay} rates for a parallel (tangential)
and perpendicular (radial) dipole orientation with 
respect to the particle surface
have been determined as (cf. Eqs. (9) and (10) of Ref. \cite{CGH})
\begin{eqnarray}
\frac{\Gamma_\parallel}{\Gamma_0} &=& 1 +\frac{3k^3}{2\pi}
\,\mbox{Im }\left[
\alpha(\omega)\exp(2ikz)
\left( \frac{1}{(kz)^6} + \frac{2}{i (kz)^5} -\frac{1}{(kz)^4} 
\right)\right],
\nonumber \\
\frac{\Gamma_\perp}{\Gamma_0} &=& 1 +\frac{3k^3}{8\pi}
\,\mbox{Im }\left[
\alpha(\omega)\exp(2ikz)\left(\frac{1}{(kz)^6}+ \frac{2}{i(kz)^5} 
           - \frac{3}{(kz)^4} \right.\right.
\nonumber\\
&& \left.\left.  - \frac{2}{i (kz)^3} 
                  + \frac{1}{(kz)^2}   \right)\right],
\label{cr:tot}
\end{eqnarray}
where $\Gamma_0$ is the (radiative) decay rate in free space, 
$z$ is the distance to the center of a
nanoparticle and $k$ stands for a wave vector. 

As in Ref. \cite{CGH}, let us focus on the distance dependence of 
the various rates of a dipole emitter at a proximity 
to Ag spherical nanoparticle (AgNP) with radius of $5$ nm. 
Two different emission wavelengths of the dipole 
emitter will be considered: $354$ nm and $612$ nm. 
The dielectric constant of AgNP at 
the respective emission wavelengths is taken to be
$\varepsilon(354 \mbox{ nm})= -2.03 + 0.6 i$ and 
$\varepsilon(612 \mbox{ nm})= -15.04 + 1.02 i$ \cite{CGH,Hand}.
For the sake of simplicity it will be  assumed throughout that the host 
medium is air. 
Already a first glance at the fluorescence decay rates
in Figs. \ref{qm612} and \ref{qm354} suffices 
to appreciate marked differences between the rates obtained 
using the 
approximate analytical approach of Carminati et al
and the exact results \cite{Rup,AMap,AMcl,Ch,Chew,KLG2}. 
The converged exact rates in Figs. \ref{qm612} and \ref{qm354} 
were obtained using the freely available code in CHEWFS \cite{AMcd}
(see Appendix \ref{app:ch} for a brief description of the code)
and comprise contributions of the induced multipoles with $l$
up to $l_{max}=50$.
Note in passing that the rates labeled as ``exact" in Figs. 1 and 2 
of Ref. \cite{CGH} are in fact those given by Eqs. (\ref{cr:tot}). 
The latter correspond to a situation when all the induced
higher order ($l>1$) multipole contributions to the 
real exact decay rates but the dipole one ($l=1$) are neglected. 
This is confirmed by Figs. \ref{qm612} and \ref{qm354},
wherein the {\em nonconverged} 
calculations of the exact decay rates with a cut-off at $l=1$
are shown to coincide with the approximate rates.
Although the authors of Ref. \cite{CGH} warned at one occasion  
that the validity of their expressions (\ref{cr:tot}) 
requires the distance $z$ to remain larger than a few radii 
of the MNP, the warning remained largely ignored.
Indeed, had the warning been taken seriously, 
any discussion of the apparent quantum yield as in Fig. 3 
of Ref. \cite{CGH} should have been
avoided in the dipole approximation: to address the 
apparent quantum yield and the question
of fluorescence quenching at short distances, one is required to 
calculate various rates for the very forbidden distances $z$ 
within a few radii of the MNP. In this work we discuss in depth 
the following issues:

\begin{enumerate}

\item The derivation of the particle polarizability 
$\alpha$ in \cite{CGH} misses a dynamic depolarization
term. 

\item Even if both the radiation reaction correction and 
dynamic depolarization are included in $\alpha$, 
there is, for the examples studied in \cite{CGH}, essentially 
no difference with regard to the corresponding rates 
obtained by substituting 
the static Rayleigh polarizability $\alpha_R$ in 
the formulas of \cite{CGH}.

\item  For the examples studied in \cite{CGH}, 
the dipole-dipole coupling cannot describe 
satisfactorily the non-radiative rates
even if the emitter is as far as three particle 
radii from the particle surface.

\item The familiar Gersten and Nitzan quasistatic theory  \cite{GN} 
yields non-radiative rates 
that agree much better with the exact rates than 
those of Ref. \cite{CGH}.

\item Because the validity of expressions (\ref{cr:tot}) 
requires the distance $z$ to remain larger than a few radii 
of the nanoparticle, the dipole approximation of Ref. \cite{CGH} 
cannot be applied to describe quenching \cite{PMS,PP,NPW,KKU}. 
The dipole-dipole approximation can only yield an upper bound 
on the apparent quantum yield.

\item All the rates are affected by the size corrections 
to the metal dielectric function \cite{KrF,BCT,AMjc}.

\end{enumerate}

The paper is organized as follows. Sec. \ref{sec:hier} introduces 
a hierarchy of approximations for the particle polarizability and 
studies the effect of a particular choice of the polarizability 
on the resulting rates. 
Although the results of Ref. \cite{CGH} have just been shown to correspond 
to the {\em nonconverged} calculations of the exact decay rates with 
a cut-off at $l=1$ (see Figs. \ref{qm612} and \ref{qm354}), 
the behaviour of the radiative rates will be shown to be different
from that of the non-radiative rates in that the $l=1$ approximation
to the radiative rates is essentially exact in the studied situation. 
The latter point is the subject of Sec. \ref{sec:rr}. 
The remaining points such as a comparison with 
the Gersten and Nitzan theory \cite{GN}, the distance dependence of 
non-radiative rates, some points regarding 
the calculation of the power $P_{abs}$ absorbed inside the particle,
and the influence of the size correction to the 
bulk dielectric function on the rates
are discussed in Sec. \ref{sec:disc}. 
We then conclude with Sec. \ref{sec:conc}.
A brief description of the freely available 
codes, which were used to generate
the results of this paper, is  provided 
in Appendix \ref{app:ch}.

\section{A hierarchy of approximations for the particle polarizability and
corresponding rates}
\label{sec:hier}
In Ref. \cite{CGH} the particle polarizability was given by
\begin{equation}
\alpha_C(\omega)  =  \frac{\alpha_R(\omega)}
              {1  - i \frac{k^3}{6\pi}\,\alpha_R(\omega)},
\label{alcr}
\end{equation}
with $\alpha_R$ (denoted as $\alpha_0$ in Ref. \cite{CGH}) 
being the familiar static Rayleigh polarizability of a sphere,
\begin{equation}
\alpha_R(\omega)= 4\pi a^3\, \frac{\varepsilon(\omega)-1}
                                    {\varepsilon(\omega)+2},
\label{sphrp}
\end{equation}
where $a$ and $\varepsilon(\omega)$ are the radius and the 
dielectric constant of the particle, respectively.
The $k^3$-dependent term is the familiar {\em radiative reaction correction} 
emphasized by earlier Ref. \cite{WGL}. 
The radiative reaction correction applies to any 
oscillating dipole, be it an elementary 
molecular dipole or the dipole induced on a small (nano)particle,
and results from the {\em Abraham-Lorentz equation}
(see, e.g., Sec. 16.2 and Eqs. (16.8-9) of Ref. \cite{J}).
However, the exact polarizability for a sphere is
obtained from  Mie's solution as 
\begin{equation}
\alpha_{Mie} =  - i\frac{6\pi}{k^3}\, T_{E1}, 
\label{tmtmie}
\end{equation}
where $T_{E1}$ is the electric dipole T-matrix element 
(the latter corresponds to the minus of the electric dipole Mie  
coefficients as given by Bohren and Huffman \cite{BH}). 
Upon introducing the size parameter $x=ka$, one finds in the 
long-wavelength limit \cite{AMdp}
\begin{eqnarray}
\alpha_{Mie} & \sim & \alpha_{Mie;as} =  4\pi a^3\,
\frac{ \varepsilon-1 }
             { \varepsilon + 2 - (6\varepsilon-12) \, \frac{x^2}{10} 
                        - i   \frac{2 x^3}{3}\, (\varepsilon-1)}
\nonumber\\
&=&  
\frac{\alpha_R(\omega)}
             { 1 -  \frac{3k^2}{20\pi a} \, 
        \frac{\varepsilon-2}{\varepsilon -1} \, \alpha_R(\omega) 
                      - i   \frac{k^3}{6\pi}\, \alpha_R(\omega)} \cdot
\label{tmtalin}
\end{eqnarray}
To this end note that each of the polarizabilities can be parametrized 
as 
\begin{equation}
\alpha = \frac{ \alpha_R }
             { 1 -  A_2 \, \frac{k^2}{4\pi a} \, \alpha_R 
                      - i A_3\,  \frac{k^3}{6\pi}\, \alpha_R} \cdot
\label{polp}
\end{equation}
Here $A_2=A_2(\varepsilon)$ ($A_2\equiv 0$ for $\alpha_R$ and $\alpha_C$) 
is in general a function of $\varepsilon$, and hence 
a {\em complex} function of frequency, 
whereas $A_3$ ($A_3=0$ for $\alpha_R$; 
$A_3= 1$ for $\alpha_C$ and $\alpha_{Mie;as}$) 
is a {\em real} numerical constant.
The special case $A_2=A_3=1$ of the above parametrization 
corresponds to the approximate Meier and  Wokaun 
polarizability $\alpha_{MW}$ \cite{MeW,AMdp}. 
The polarizabilities $\alpha_R$, $\alpha_C$, $\alpha_{MW}$, and
$\alpha_{Mie;as}$ 
in this order approximate  the particle polarizability $\alpha_{Mie}$ 
with an increasing precision. For instance, when $\alpha_C$ is compared 
against $\alpha_{MW}$ or $\alpha_{Mie;as}$,
one immediately notices a missing $k^2$-term in the denominator of 
$\alpha_C$. The missing term is the 
so-called {\em dynamic depolarization} term \cite{MeW,AMdp}.
For a given constant homogeneous external field, 
the Meier and Wokaun polarizability
$\alpha_{MW}$ presumes a constant internal field within the particle.
In contrast to $\alpha_{MW}$, $\alpha_{Mie;as}$ takes 
implicitly into account an order $x^2$ deviation from the constant 
internal field \cite{AMdp}.

%
%
\begin{center}
TABLE 1. The values of various polarizabilities 
$\alpha = $(Re $\alpha$, Im $\alpha$) of AgNP for \\
$\lambda=354$ nm and $\lambda=612$ nm. 
\vspace*{0.5cm} \\
\begin{tabular} {|c|c|c|} \hline
 &  $\lambda=354$ nm &  $\lambda=612$ nm \\
  \hline 
 $\alpha_R/(4\pi a^3)$ &  (1.249, 4.988) &  (1.229, 0.0179)
\\ 
$\alpha_C/(4\pi a^3)$ &  (1.244, 4.977) &  (1.229, 0.0180)
\\ 
$\alpha_{MW}/(4\pi a^3)$ &  (1.055, 5.068)  &  (1.233, 0.0181)
\\ 
$\alpha_{Mie;as}/(4\pi a^3)$ &  (1.119, 5.036) &  (1.231, 0.0181)
\\ 
 $\alpha_{Mie}/(4\pi a^3)$  &  (1.093, 5.042)  &  (1.231, 0.0181)
 \\ 
 \hline
\end{tabular}
\end{center}
\vspace*{0.5cm} 
%
%
\noindent
As shown in Ref. \cite{AMdp}, the dynamic depolarization term can also
be derived from the very same Green's function approach of Ref. \cite{CGH},
the fact that has eluded the authors of Ref. \cite{CGH}.
Without the dynamic depolarization term, the polarizability $\alpha_{C}$
does not take into account any size dependence of the dipolar 
localized surface plasmon resonance (LSPR) position.
The size dependence is determined, up to the order of $x^2$, 
by the following equation for the real part of 
$\varepsilon= \varepsilon' + i\varepsilon''$ (see Sec. 12.1.1 of 
Ref.~\cite{BH}):
\begin{equation}
\varepsilon' \approx - 2 - \frac{12 x^2}{5}\cdot
\label{bhexp}
\end{equation}
Although the dynamic depolarization term would 
cause for $a\gtrsim 40$ nm an appreciable difference of the rates obtained 
with $\alpha_{Mie;as}$ or $\alpha_{Mie}$ compared to those with
$\alpha_C$, with the radius of AgNP of mere $5$ nm and $x < 0.09$, 
there is, in the present case, 
essentially no difference as to what polarizability is
substituted into the approximate rate equations of Ref. \cite{CGH}.
The latter point is illustrated by Table 1 that lists the values
of various polarizabilities, which have been introduced so far.
Therefore, the difference in the fluorescence decay rates obtained 
using the 
approximate analytical approach of Carminati et al \cite{CGH}
with regard to the exact results \cite{Rup,AMap,AMcl,Ch,Chew,KLG2}
(see Figs. \ref{qm612} and \ref{qm354}) cannot be the result
of a wrong choice of polarizability. 

To this end note that the implications of the unitary bound
on the physical S matrix,
\begin{equation}
0 \leq SS^\dagger \leq 1.
\label{untbnd}
\end{equation}
Given the parametrization $S=1+2iT$ of the S matrix, 
the unitarity bound implies for the physical T matrix and the 
polarizability 
\begin{equation}
-\frac{1}{4} - |T|^2 \leq \mbox{Re }T \leq - |T|^2 \leq 0,
   \hspace*{0.4cm}
\frac{k^3}{6\pi}\, |\alpha|^2 \leq \mbox{Im }\alpha \leq 
                    \frac{k^3}{6\pi}\, |\alpha|^2 + \frac{3 \pi}{2k^3},
\label{talbndsi}
\end{equation}
where the second bound follows from the first one 
on substituting $ik^3\alpha/(6\pi)$ for $T$.
Because $\mbox{Im }\alpha_R \equiv 0$ for a real $\varepsilon$, 
the familiar static Rayleigh polarizability $\alpha_R$
violates the above physical bounds. The violation is reflected
by the fact that $\sigma_{tot} \neq \sigma_{sca} + \sigma_{abs}$, where
$\sigma_{tot}$, $\sigma_{sca}$, $\sigma_{abs}$ are the total, scattering, 
and absorption cross-sections. Nevertheless, 
since $\sigma_{tot}$ is in the present case dominated by $\sigma_{abs}$,
one could in the present case calculate the rates 
with the familiar static Rayleigh polarizability $\alpha_R$ 
without ever noticing any difference with respect to 
the results obtained by means of $\alpha_C$
(see Figs. \ref{rrad612} to \ref{nrad354}).

\section{Radiative and non-radiative rates}
\label{sec:rr}
The non-radiative rates are derived from the power $P_{abs}$ 
absorbed inside the particle \cite{Rup,AMap,CGH,KLG2}. 
Carminati et al \cite{CGH} derived 
the following approximate expressions for the rates 
(see Eqs. (15) and (16) of Ref. \cite{CGH}):
\begin{eqnarray}
\frac{\Gamma_{nr;\parallel}}{\Gamma_0} &=& \frac{3 k^3}{2\pi} \left[
\mbox{Im }[\alpha(\omega)]  - \frac{k^3}{6\pi} |\alpha(\omega)|^2 \right]
\left[ \frac{1}{(kz)^6}+  \frac{1}{(kz)^4} \right],
\nonumber
\\
\frac{\Gamma_{nr;\perp}}{\Gamma_0} &=& \frac{3 k^3}{8\pi} \left[
\mbox{Im }[\alpha(\omega)]  - \frac{k^3}{6\pi} |\alpha(\omega)|^2 \right]
\left[ \frac{1}{(kz)^6} -  \frac{1}{(kz)^4} + \frac{1}{(kz)^2} \right].
\label{crnr}
\end{eqnarray}
The respective radiative rates were then derived in Ref. \cite{CGH} as a 
difference of the corresponding total decay rates in the 
short distance limit $ka<kz\ll 1$ (Eqs. (11),(12) in Ref. \cite{CGH})
and the non-radiative rates (\ref{crnr}).
The resulting approximate expressions for 
the radiative rates were (Eqs. (17) and (18) of Ref. \cite{CGH}):
\begin{eqnarray}
\frac{\Gamma_{r;\parallel}}{\Gamma_0} &=& 1 +\frac{k^6}{4\pi^2}
|\alpha(\omega)|^2 \,\left[ \frac{1}{(kz)^6}+  \frac{1}{(kz)^4}
\right] + \frac{k^3}{\pi}\, \mbox{Re } [\alpha(\omega)] \, \frac{1}{ (kz)^3},
\nonumber
\\
\frac{\Gamma_{r;\perp}}{\Gamma_0} &=&  1 +\frac{k^6}{16\pi^2}
|\alpha(\omega)|^2 \,\left[ \frac{1}{(kz)^6} -  \frac{1}{(kz)^4}
\right] - \frac{k^3}{2\pi}\, \mbox{Re } [\alpha(\omega)] \, \frac{1}{ (kz)^3}\cdot
\label{crrr}
\end{eqnarray}
It can be verified 
that the above formulae of Ref. \cite{CGH} hold
irrespective which polarizability of Sec. \ref{sec:hier} is used.
A necessary and sufficient condition is to show
that the power $P_{abs}$ absorbed inside the particle 
remains to be given by the formula (see Eq. (13) of Ref. \cite{CGH})
\begin{equation}
P_{abs} = \frac{\omega\varepsilon_0}{2 } \,\left( \mbox{Im } \alpha  
        -  \frac{k^3}{6 \pi}\, |\alpha|^2\right)  |{\bf E}_{exc}|^2.
\label{my13sc}
\end{equation}
The latter formula can easily be verified to follow from
the more general expression
\begin{equation}
P_{abs} = \frac{\omega\varepsilon_0}{2 k} \, \sigma_{abs} |{\bf E}_{exc}|^2,
\label{my13}
\end{equation}
where $\sigma_{abs}$ is the absorption cross-section of a MNP. 
Indeed, provided that one factorizes the electric dipole element of 
the T-matrix (in the SI units) as
$T_{E1}= i\frac{k^3}{6\pi}\, \alpha$ [cf Eq. (\ref{tmtmie})], 
the total (extinction), scattering, and  absorption
cross sections can be expressed as
\begin{eqnarray}
\sigma_{tot} &\approx &  -\frac{6\pi}{k^2}\, \mbox{Re } T_{E1} 
        =   k\, \mbox{Im}\, \alpha,
\label{sgtot}
\\
\sigma_{sca} &\approx & \frac{6\pi}{k^2}\, |T_{E1}|^2 
             = \frac{k^4}{6\pi}\, |\alpha|^2,
\label{scad}
\\
\sigma_{abs}  &\approx & \frac{3\pi}{2k^2}\, \left( 1- |1+2T_{E1}|^2\right)
             = k\, \left( \mbox{Im } \alpha  -  
               \frac{k^3}{6 \pi}\, |\alpha|^2\right).
\label{sgabssi}
\end{eqnarray}
Consequently, upon substituting (\ref{sgabssi}) into (\ref{my13})
one recovers the formula (\ref{my13sc}), 
irrespective of the choice of the polarizability of Sec. \ref{sec:hier}.

A comparison of the approximate and exact radiative rates 
is shown in Figs. \ref{rrad612} and \ref{rrad354}.
The results show clearly that all rates coalesce to a single line, 
irrespective if obtained in the dipole-dipole approximation 
with $\alpha_R$, $\alpha_C$, $\alpha_{Mie;as}$, or
 $\alpha_{Mie}$, or as the exact radiative rates with a cut-off 
imposed at either $l=1$ or $l=50$. Thus the induced
multipoles of the order $l\geq 2$ have in the present case hardly any effect 
on the radiative rates, because
the exact radiative rates are very precisely
approximated by taking into account solely the dipole-dipole interaction. 
The latter means that the conclusions of Carminati et al \cite{CGH} 
regarding the distance dependence of the radiative 
decay rate are correct. The distance dependence 
is chiefly dominated by a $z^{-3}$ dependence, 
with a $z^{-6}$ dependence being visible at plasmon resonance. 
On the other hand the results are hardly surprising and appear
to have been anticipated by Gersten and Nitzan, who also approximate
the exact radiative rates by taking into account solely 
the dipole-dipole interaction (see Eqs. (B.18') and 
(B.43') of Mathematical Appendices to Ref. \cite{GN}).

An analogous comparison for the non-radiative rates 
is shown in Figs. \ref{nrad612} and \ref{nrad354}.
Because the leading contribution to the non-radiative rates (\ref{crnr})
is proportional to $\mbox{Im }\alpha$, and $\alpha$ reaches a
maximum at the LSPR, the non-radiative decay 
is strongly enhanced when the emitter radiates at the 
plasmon-resonance frequency of the nanoparticle. 
Again all the approximate dipole-dipole interaction results, irrespective if 
obtained with $\alpha_R$, $\alpha_C$, $\alpha_{Mie;as}$, or $\alpha_{Mie}$, 
together with the exact radiative rates obtained 
with a cut-off imposed at $l=1$ coalesce to a single line.
However, contrary to the radiative rates shown in Figs. \ref{rrad612} and 
\ref{rrad354}, the non-radiative rates cannot be approximated by 
taking into account solely the dipole-dipole interaction.
Clearly, as demonstrated by the exact non-radiative rates with 
the cut-offs imposed at $l=2$, $l=4$, and $l=50$, respectively,
the contribution of induced multipoles of the order $l\geq 2$ has 
a pronounced  effect on the non-radiative rates.
The latter shows up in the distance dependence of the non-radiative
rate. This point is further elaborated in 
Sec. \ref{ssec:dd}. 
Because the exact $\Gamma_{nr}$ is always higher than the approximate one,
whereas the exact and approximate $\Gamma_{r}$ are essentially equal, 
the dipole-dipole approximation of Ref. \cite{CGH}
can only provide an upper bound on the apparent 
quantum yield $Y$ that is defined as the ratio 
$Y=\Gamma_r/(\Gamma_r+\Gamma_{nr})$.
The different distance behavior of the radiative 
and non-radiative rates implies that the apparent quantum yield 
always vanishes at short distance, resulting in a
quenching of the emission.

The results for the exact radiative and non-radiative 
rates shown in the figures were obtained using the freely available 
code CHEW \cite{AMcd1} (see Appendix \ref{app:ch} for a brief 
description of the code). The exact rates comprise contributions 
of the induced multipoles with $l$ up to $l_{max}=50$.
The exact radiative rates are derived from the total radiated power.
The latter is calculated by the code CHEW by means of 
Eq. (108) of \cite{AMap}, where, in the notation of Ref. \cite{AMap},
the functions $f_{\gamma l}$ are given as linear combinations of spherical
functions (Eqs. (69), (72) of \cite{AMap}) taken at the dipole position,
with the coordinate origin located at the sphere center.
[In the special case of a dipole outside a homogeneous sphere, 
the coefficients $Q_{\gamma l}$ (Eq. (72) of \cite{AMap}) that 
enter the definition of $f_{\gamma l}$ (Eq. (69) of \cite{AMap})  reduce 
to the Mie's expansion coefficients (cf. Eqs. (32), (33) of  \cite{AMap}).]
The exact power $P_{abs}$ absorbed inside the particle 
is calculated by the code CHEW 
by means of Eq. (119) taken in combination with 
Eqs. (116) and (120) of \cite{AMap}.
In the special case of a dipole outside a homogeneous sphere one 
has, in the notation of Ref. \cite{AMap}, $N=n=1$, and
the coefficients $a_{\gamma l}$ and $b_{\gamma l}$ in the integrands of 
the respective integrals $I^{(s)}_{\gamma l}$ in
Eq. (116) of \cite{AMap} reduce to
\begin{equation}
a_{\gamma l} = 1 / {\cal T}_{11;\gamma l}(2)
= 1 / T^+_{11;\gamma l}(1),
\hspace*{1.2cm}
b_{\gamma l} = 0.
\label{acdrelcsc}
\end{equation}

It is worth to emphasize that the exact rates shown in the
figures were all 
calculated independently: (i) the fluorescence decay rates 
have been determined in CHEWFS from the imaginary part of a 
relevant dyadics (cf. Eq. (1) of Ref. \cite{CGH}
Eq. (135) of Ref.  \cite{AMap}), the radiative rates have been 
determined in CHEW from the total radiated power of 
a dipole escaping to infinity (Eqs. (108) and (124) of 
Ref. \cite{AMap}), and the non-radiative 
rates have been determined in CHEW from the total Ohmic loss 
of dipole radiation (see Eqs. (111), (115), (116), (133) 
of Ref. \cite{AMap}). One can verify that the sum of the
radiative and non-radiative rates as obtained using CHEW 
coincides with the total decay rate as obtained using CHEWFS.

\section{Discussion}
\label{sec:disc}

\subsection{A comparison with the Gersten and 
Nitzan rates}
In the special case of a dipole outside a homogeneous sphere,
the functional form of the respective exact expressions for the 
total radiated power [Eq. (108) of \cite{AMap} with 
$f_{\gamma l}=j_l(kr_d) + \kappa_{\gamma l} h_l^{(1)}(kr_d)$, where
$\kappa_{\gamma l}$ are the Mie's coefficients ($a_l$ for $\gamma=M$ and 
$b_l$ for $\gamma=M$; cf Eqs. (69), (32), and (34) of \cite{AMap})] and 
for the power $P_{abs}$ absorbed inside the particle [Eq. (119) 
of \cite{AMap} with $d_{\gamma l}=h_l^{(1)}(kr_d)$]
look quite similar. However, whereas in Eq. (108) of \cite{AMap} 
the Hankel functions are with increasing
$l$ multiplied by rapidly decreasing Mie's coefficients,  
in Eq. (119) of \cite{AMap} 
the Hankel functions are multiplied with increasing
$l$ by  much slowly decreasing values
of the integrals $I^{(s)}_{\gamma l}$ (Eqs. (116) of \cite{AMap}) 
taken over the particle volume.
The latter explains, which
is also an intrinsic feature of
the approximate quasistatic Gersten and Nitzan theory \cite{GN},
that in the small particle limit
(i) $l=1$ is enough to describe the radiative 
rates (see Eqs. (B.18') and 
(B.43') of Mathematical Appendices to Ref. \cite{GN}), 
whereas, in principle, (ii) an infinite series in $l$ is required
to describe the non-radiative rates 
(see Eqs. (B.24') and (B.45') of Mathematical Appendices to Ref. \cite{GN}).
This is also the reason why the {\em non-radiative} rates 
calculated by the Gersten and Nitzan theory agree much better
with the exact electrodynamic non-radiative rates as confirmed
by Figs. \ref{gnnrex612} and \ref{gnnrex354}.
In arriving at the results, the same formulas as Eqs. (3) and (4) of 
Ref. \cite{SRW} were used with the respective terms $|1-\Delta|^2$,
where $\Delta(\perp)$ and $\Delta(\parallel)$ are the Gersten and Nitzan 
image enhancement factors, being essentially equal to one.
(In the present case, the ratios $\Delta/\Gamma_{r,int}$,
with $\Gamma_{r,int}$ denoting the intrinsic vacuum radiative rate,
are of the order of $10^{-10}$.) That is analogous to Ref. \cite{KLG}, 
where $|1-\Delta|^2\approx 1$ is implicit upon comparing Eqs. (5)
and (6) with the exact formulas (B.18') and (B.43') of Mathematical 
Appendices to Ref. \cite{GN}. 
The above conclusions, i.e. that the molecular dipole  excites 
all the multipoles of the
sphere, but only the dipole radiates, whereas higher-order multipole terms 
gives rise to dissipation, have  been confirmed within the quasistatic
approximation also by other researcher \cite{FoW1,Mtu,BhN}.

The Gersten and Nitzan results are supported 
by experiment \cite{SRW}. A mismatch between the theoretical and the 
measured non-radiative rates 
reported in earlier work \cite{Dulk,DRK} may partly be attributed 
to nonlocal effects \cite{VLe} and, partly, to the fact
that a point dipole may be a too crude approximation for fluorophores
but remains rather good for phosphors (the latter have a dipole strength 
that is reduced by orders of magnitude 
compared to that of fluorophores \cite{SRW}).
The nonlocal effects, in general, lead to significantly 
{\em greater} fluorescence rates and {\em smaller} non-radiative 
decay rates for the admolecules \cite{VLe}.
Eventually, Sec. \ref{sec-scdf} below demonstrates that  
the theoretical values of non-radiative rates 
for fluorescence near a LSPR
can be further decreased toward the 
measured non-radiative rates \cite{Dulk,DRK} by acounting for 
size corrected dielectric function.

Although the approximate quasistatic Gersten and Nitzan theory \cite{GN}
has been shown to be superior to the dipole-dipole approximation of 
Ref. \cite{CGH}, the use of the Gersten and Nitzan theory \cite{GN} is also not 
without limitations \cite{KLG}. For instance, in contrast 
to the exact Chew expressions for the {\em radiative} rates \cite{Ch,Chew}, 
the theory of Gersten and Nitzan \cite{GN} does not yield the correct 
asymptotic results for a flat surface in the limit $a \rightarrow \infty$ 
with a fixed value of the separation $d$ of the dipole from the surface of 
a sphere. Moreover, as emphasized by Kim et al \cite{KLG},
in addition to the conditions $a \ll\lambda$ and $d\ll\lambda$, 
the value of sphere radius $a$ cannot be too large for a fixed $d$ 
if the Gersten and Nitzan model \cite{GN}, and hence any dipole-dipole
coupling model, is to be applied for the {\em radiative} rates. 
Eventually, contrary to the exact electrodynamic results,
the ratios $\Gamma_{nr}/\Gamma_{0}$ and $\Gamma_{r}/\Gamma_{0}$, 
depend on the intrinsic vacuum radiative rate, because 
both image enhancement factors 
$\Delta(\perp)$ and $\Delta(\parallel)$ of the Gersten and Nitzan model 
depend on $\Gamma_{r,int}$ \cite{GN}.

\subsection{The distance dependence of non-radiative rates}
\label{ssec:dd}
The approximate non-radiative rates (\ref{crnr}) gave an impression
that their short distance dependence is, like in the F\"{o}rster 
energy transfer, characterized by a $z^{-6}$-like dependence.
However, it has been just shown (see Figs. \ref{nrad612} and \ref{nrad354}) 
that a dipole-dipole coupling is not appropriate to describe the 
non-radiative rates of a dipole emitter in a proximity of a MNP.
Taking for example $\Gamma_{nr;\parallel}$, 
the exact $\Gamma_{nr;\parallel}$ for the 
emission wavelength of $612$ nm shown in Fig. \ref{nrad612}
is at the distances of $5$ nm, $10$ nm, and $15$ nm
from the AgNP surface by $101\%$, $33\%$, and $17\%$ higher than 
the approximate dipole-dipole contribution of Ref. \cite{CGH}.
Similarly, the exact $\Gamma_{nr;\parallel}$ for the 
emission wavelength of $354$ nm shown in Fig. \ref{nrad354} 
is at the distances of $5$ nm, $10$ nm, and $15$ nm
from the AgNP surface by $56\%$, $20\%$, and $11\%$ higher than 
the approximate dipole-dipole contribution of Ref. \cite{CGH}.
Because of the importance of the higher-order ($l>1$) multipoles contribution,
which decay as $z^{-(2l+4)}$,
it makes rather sense to consider the dependence of the 
non-radiative rates
on the separation $d$ of the dipole from the surface of 
a sphere \cite{FoW1,Dlkt}. 
Depending on the particle radius, the non-radiative rates
can decay between $\approx d^{-5}$ and $d^{-3}$, wherein
a near $d^{-3}$ decay is already observed for the particle
radii $\gtrsim 10$ nm (see Figs. 16 and 17 in Sec. 3.6 of Ref. \cite{FoW1} and 
Fig. 2.15 of Ref. \cite{Dlkt}).
A slower than $d^{-6}$ dependence of the non-radiative rates
is valid even for particles as small as $1$ nm 
in diameter \cite{Dlkt}. 
Upon taking into account the results by Dulkeith
\cite{Dlkt}, one can conclude that even for MNPs with the radius as small 
as $0.5$ nm and the emitter as far as the distance of $3a$ from 
the particle surface, the emitter and the MNP do not interact 
exclusively through dipole-dipole coupling.

A full quantum-mechanical calculation for a {\em spherical} 
metal surface obtained within the time-dependent local density 
approximation (TDLDA) by Ekardt and Penzar \cite{EkP} 
essentially confirms the above picture.  
Formally similar to the classical electrodynamic case \cite{Rup,AMap},
the quantum-mechanical non-radiative rates (see Eq. (5) in Ref. \cite{EkP}) 
were given as an infinite series over $l$ of the terms proportional 
to $(\mbox{Im }\alpha_l)/z^{2l+4}$, where $\alpha_l$ is the $l$-pole 
polarizability of the sphere. 
The difference with the classical case is in that
$\alpha_l(\omega)$ is the quantum-mechanical {\em dynamical} 
$l$-pole polarizability 
calculated within the TDLDA. 
Because each term in the series has a different 
distance dependence, the overall distance dependence of 
the non-radiative rates cannot be uniquely determined, simply 
because everything is mixed up \cite{EkP}.
Note in passing that the TDLDA results incorporate 
the so-called nonlocal dielectric response \cite{VLe,AgaN,Leu}
without any approximation.

Unfortunately, a detailed comparison of the classical electrodynamic results
\cite{Rup,AMap} against the full quantum-mechanical calculation remains
to be an open question. An obstacle in comparing quantum-mechanical TDLDA
with the classical electrodynamics is that the former works
with an electron density parameter \cite{EkP}, from which there is still
some distance to calculate the dielectric function of say silver,
which would then serve as the input parameter for the
classical electrodynamic calculations.
Thus at present it is not possible to answer conclusively
at which distances from a MNP surface the classical results 
can no longer be applied.

\subsection{The power $P_{abs}$ absorbed inside the particle}
\label{sc:p}
It turns out that it is more reliable to calculate
the power $P_{abs}$ absorbed inside the particle by using 
the formula (\ref{my13}). On the other hand, 
a direct integration approach 
of Carminati et al \cite{CGH} leads to a contradiction.
Indeed, according to the definition of the polarizability,
the field inside the particle, ${\bf E}_{in}$, is related to the external 
field ${\bf E}_{exc}$ acting on the particle as follows:
\begin{equation}
{\bf E}_{in} = \frac{3\alpha}{4\pi a^3 (\varepsilon-1)}\, {\bf E}_{exc}.
\label{inex}
\end{equation}
On using the parametrization (\ref{polp}) of the polarizability,
\begin{equation}
{\bf E}_{in} =    \frac{3}{(\varepsilon+ 2)}
       \frac{{\bf E}_{exc}}{\left( 1 -  A_2\, \frac{k^2}{4\pi a} \, \alpha_R 
                      - i A_3\,  \frac{k^3}{6\pi}\, \alpha_R\right)}\cdot
\label{polp1}
\end{equation}
When this expression is substituted into the formula 
(B.1) of Ref. \cite{CGH},
\begin{equation}
P_{abs} = \frac{4}{3}\, \pi a^3 \, \frac{\omega\varepsilon_0}{2}\, 
                  \varepsilon''\, |{\bf E}_{in}|^2,
\label{b1}
\end{equation}
one arrives at
\begin{equation}
P_{abs} = 4\pi a^3\,\frac{\omega\varepsilon_0}{2}\,  
         \frac{3\, \varepsilon''}{|\varepsilon+ 2|^2}\, 
  \frac{|{\bf E}_{exc}|^2}{\left| 1 -  A_2\, \frac{k^2}{4\pi a} \, \alpha_R 
                      - i A_3\,  \frac{k^3}{6\pi}\, \alpha_R\right|^2}\cdot
\end{equation}
With the help of the identity (B.3) of Ref. \cite{CGH},
\begin{equation}
4\pi a^3\,\frac{3\, \varepsilon''}{|\varepsilon +2|^2} =
4\pi a^3\, \mbox{Im } \left( \frac{\varepsilon -1}
  {\varepsilon+2}\right)=\mbox{Im } \alpha_R,
\label{b3}
\end{equation}
and upon comparing with (cf. Eq. (B.5) of Ref. \cite{CGH})
\begin{equation}
\mbox{Im } \alpha = \frac{(\mbox{Im } \alpha_R)}
   {\left|1 - A_2\, \frac{k^2}{4\pi a}\,\alpha_R
         - i A_3\, \frac{k^3}{6\pi}\,\alpha_R\right|^2}
+ 
\left[ (\mbox{Im }A_2)\, \frac{k^2}{4\pi a} 
    + A_3\, \frac{k^3}{6\pi}\right] \, |\alpha|^2,
\label{alb5}
\end{equation}
that follows from the parametrization (\ref{polp}),
one eventually arrives at
\begin{equation}
P_{abs} = 
\frac{\omega\varepsilon_0}{2}\,  
\left\{ \mbox{Im } \alpha -
\left[ (\mbox{Im }A_2)\, \frac{k^2}{4\pi a} 
    + A_3\, \frac{k^3}{6\pi}\right] \, |\alpha|^2
\right\}\,|{\bf E}_{exc}|^2.
\label{pabs}
\end{equation}
Actually $\mbox{Im }A_2\neq 0$ only for $\alpha_{Mie;as}$, in which case
\begin{equation}
\mbox{Im }A_2 =\frac{3}{5}\frac{\varepsilon''}{|\varepsilon-1|^2},
\end{equation}
whereas $\mbox{Im }A_2\equiv 0$ for $\alpha_R$, $\alpha_C$, and $\alpha_{MW}$.
Upon comparing the two expressions (\ref{my13sc}) and 
(\ref{pabs}) for the absorbed power $P_{abs}$ one finds
that, in the case of $\alpha_{Mie;as}$, they differ in the term proportional 
to $\mbox{Im }A_2$. The result (\ref{pabs}) would
imply a modification of the non-radiative and radiative rate formulae 
of Ref. \cite{CGH}. However, the latter difference of the two expressions 
for $P_{abs}$ appears to be a consequence of that Eq. (\ref{b1})
is only an approximation to
\begin{equation}
P_{abs} =  \frac{\omega\varepsilon_0}{2}\,  
         \varepsilon''\, \int_V |{\bf E}_{in}|^2 \,dV,
\label{b1f}
\end{equation}
where the integral is over the volume of the particle. 
Eqs. (\ref{my13sc}) and (\ref{b1f}) are equivalent.
However, the hypothesis that, for a given constant homogeneous external 
field, the internal field inside the 
particle is a constant, and which leads to Eq. (\ref{b1}), introduces
an order $x^2$ deviation from the exact result. 
Although at a first approximation
the internal field inside the  particle can indeed be assumed
to be a constant, this is of course not true exactly \cite{Stev2}.
It has been shown that the very form of the dynamic depolarization
term in $\alpha_{Mie;as}$ implicitly implies an order $x^2$ 
deviations from the constant internal field \cite{AMdp}. 
The order $x^2$ deviation from the exact result 
for $P_{abs}$ when using Eq. (\ref{b1}) is of the same order 
as the $\mbox{Im }A_2$ term that
makes the difference between Eqs. (\ref{my13sc}) and 
(\ref{pabs}). It is therefore natural to assume that
if the volume integration in Eq. (\ref{b1f}) were performed 
by taking into account the order $x^2$ deviation from the
constant internal field, one would arrive at the 
result (\ref{my13sc}) obtained directly
from the absorption cross section.

\subsection{Effect of size corrections to the bulk dielectric function}
\label{sec-scdf}
It is generally accepted that for radii of MNPs smaller than the mean
free path of conduction electrons ($42$ nm for Au and $52$ nm for Ag)
the dielectric function of free electron metals should be corrected
due to a reduced mean-free path of the electrons because of the
scattering of the electrons on the MNP surface \cite{KrF}
\begin{equation}
\varepsilon_s (\omega) = \varepsilon_{b}(\omega) - 
  \varepsilon^{D}_{b} (\omega) + \varepsilon^{D}_{sd}(\omega). 
\label{bdsc}
\end{equation}
Here, $\varepsilon_{b}$ is the bulk metal dielectric function, and 
$\varepsilon^{D}_{b}$ is the Drude dielectric function
\begin{equation} 
\varepsilon^{D}_{b} (\omega) = 1-\omega_p^2/[\omega(\omega+i\gamma)],
\label{bd}
\end{equation} 
describing the bulk metal conduction electrons 
with the plasma frequency $\omega_p$ and the bulk damping constant $\gamma$. 
The term $\varepsilon^{D}_{sd}$ differs from $\varepsilon^{D}_{b}$ in that 
$\gamma$ is replaced by the size-corrected value \cite{KrF,AMjc}
\begin{equation}
\gamma_{sd} = \gamma + A v_F/a,
\label{gmsc}
\end{equation}
where $v_F$ is the Fermi velocity and $A$ 
(not to be confused with $A_j's$ in the preceding sections)
is a phenomenological fitting parameter \cite{KrF}. 
For gold and silver the size-correction becomes appreciable
when, as in the present case, the sphere radius is $\lesssim 8$ nm  
(see for instance Fig. 4 \cite{BCT}), i.e.
less than one fifth of the mean
free path of conduction electrons.
The following parameters were employed for silver:
$v_F=1.39$ nm fs${}^{-1}$, $\omega_p=72700$ cm${}^{-1}$, 
and $\gamma=145$ cm${}^{-1}$ \cite{OLB1,Drd}.
Given the above parameters, 
$\varepsilon(354 \mbox{ nm})= -2.03 + 0.6 i$ 
would change into $\varepsilon(354 \mbox{ nm}) \approx -2.028 +0.686 i$ 
for $A=0.25$ and 
into $\varepsilon(354 \mbox{ nm}) \approx -2.008 +0.945i$ for $A=1$.
Similarly, $\varepsilon(612 \mbox{ nm})= -15.04 + 1.02 i$ 
would change into $\varepsilon(612 \mbox{ nm})= -15.022 + 1.466 i$ for $A=0.25$ 
and into $\varepsilon(612 \mbox{ nm})= -14.849 + 2.789 i$ for $A=1$.
Following experimental results
for individual gold NPs down to $a=2.5$ nm by Berciaud et al \cite{BCT},
the value of $A=0.25$ could be employed  on the grounds of the similarity 
in the band structure of silver and gold \cite{BCT}. 
Depending on the sample preparation, the value of $A=1$
may be suitable in other cases \cite{KrF}.

As shown in Figs. \ref{gnnrex612} and \ref{gnnrex354}
 by a dash-dot line, the effect of the size corrected dielectric function 
on the non-radiative rates is rather pronounced for $A=1$. 
At the wavelength of $354$ nm, the size correction 
has the effect of decreasing the non-radiative rates, whereas 
for the wavelength of $612$ nm the size correction has the opposite effect.
Although the above results may at a first glance appear counterintuitive,
they can be explained rather straightforwardly. Take for instance
the leading  $l=1$ contribution, which is well approximated 
by Eqs. (\ref{crnr}). Eqs. (\ref{crnr}) show that the leading  $l=1$ contribution
to the non-radiative decay is controlled through $\mbox{Im } \alpha$ rather
than $\varepsilon''$. According to Eq. (\ref{b3}), 
$\mbox{Im } \alpha_R$ depends on $\varepsilon''$ as $3/(\varepsilon'')^2$
near the dipolar LSPR.
Therefore, at the proximity of the LSPR, increasing $\varepsilon''$  
results in {\em smaller} values of $\mbox{Im } \alpha_R$ that 
causes the non-radiative decay to {\em decrease}. 
Thus by properly acounting for the
size corrected dielectric function at the proximity of the LSPR
may yield an additional
mechanism for decreasing the theoretical values of non-radiative rates 
toward the measured non-radiative rates \cite{Dulk,DRK}. 

On the other hand, $\varepsilon'$ is typically ten 
times larger in magnitude than $\varepsilon''$ far away from 
the LSPR. Consequently a change in $\varepsilon''$ has only a minor effect
on the denominator Eq. (\ref{b3}) and $\mbox{Im } \alpha_R$ becomes
essentially proportional to $3\varepsilon''$. Thus far away from 
the LSPR increasing $\varepsilon''$ results in {\em larger} value of
$\mbox{Im } \alpha_R$ that causes the non-radiative decay to {\em increase}.

\section{Conclusions}
\label{sec:conc}
The methods based on dipole-dipole coupling \cite{CGH,GN}
have known limitations in describing the radiative rates \cite{KLG}.
In line with earlier quasistatic results \cite{GN,FoW1,Mtu,BhN},
the present work reaffirms that the dipole-dipole coupling  
severely underestimate the non-radiative rates even
in the parameter range, where it rather satisfactorily describes
the radiative rates. 
An approximate analytical dipole-dipole coupling approach, 
such as that of Ref. \cite{CGH}, may fail without any warning even for relatively 
small particles. The dipole-dipole interaction results 
of Carminati et al \cite{CGH} cannot be improved by any generalization 
of their expression for the dipolar polarizability, either by supplying
the dynamic depolarization term or by taking the full Mie's 
expression. To correctly describe the radiative 
and non-radiative decay 
of fluorophores, chromophores, and quantum dots close to a metallic nanoparticle, 
the use of a complete theory is recommended \cite{Rup,AMap,AMcl,Ch,Chew,KLG2}.
Taking a Ag spherical nanoparticle (AgNP) with the radius of $5$ nm as an example, 
a significant contribution (between $50\%$ and $101\%$ of the total
value) of the higher-order ($l>1$) multipoles to the non-radiative rates was
found at the emitter distance of $5$ nm from the AgNP surface. 
Consequently, a dipole-dipole approximation can yield only an upper bound on the 
apparent quantum yield. In general, the higher-order ($l>1$) multipole contributions to
the non-radiative rates cannot be neglected, 
even if a particle has a radius as small as $a=0.5$ nm and 
the dipole emitter is further than $15$ nm from the particle surface.
The effect of the higher-order multipole contributions
shows up in a markedly slower than $d^{-6}$ distance dependence
of the non-radiative rates. The non-radiative rates calculated 
with the approximate quasistatic Gersten and Nitzan method, which takes into account
the contribution of higher-order ($l>1$) multipoles, were 
found to be in much better agreement with exact electrodynamic results. 
Finally, the size
corrected metal dielectric function was shown to decrease the
non-radiative rates near the dipolar surface plasmon resonance.

\section{Acknowledgments}
I thank G.-Y. Guo and R. Hammerling for a careful
reading of the manuscript and useful suggestions.

\appendix

\section{A brief description of the codes CHEW and CHEWFS
with instructions how to adapt them 
to case of a homogeneous sphere}
\label{app:ch}

The Fortran F77 codes CHEWFS \cite{AMcd}  and CHEW \cite{AMcd1}, 
which have been written 
according to the theory developed in Refs. \cite{AMap,AMcl} and
which have been free available since October 2004, are designed to
calculate various decay rates of a dipole emitter interacting 
with a multi-coated sphere with an arbitrary number of shells.
The emitter can be located either inside or outside a sphere.

CHEWFS \cite{AMcd} calculates the total decay rate and frequency shift
directly from the respective imaginary and  real parts of 
the Green's function at coinciding arguments (cf. Eqs. 
(135) and (137) of Ref. \cite{AMap}). 
CHEWFS returns files {\em dpfshiftm.dat} (containing frequency shifts) and
{\em dipqmratesm.dat} (containing total decay rates).

CHEW \cite{AMcd1} calculates the radiative and non-radiative  rates, which are
both calculated classically. The radiative rate is determined from the 
total radiated power
of a dipole escaping to infinity. The latter
is found by integrating the radial component 
of the time-averaged Poynting vector over the surface of a sphere with a 
radius going to infinity (cf. Eq. (88) of Ref. \cite{AMap}).
The output is in files {\em dipratesm.dat} and {\em dipratesv.dat}. 
The non-radiative rates are determined by CHEW from
the Ohmic losses inside absorptive shells 
(cf. Eqs. (110) and (111) of Ref. \cite{AMap}).
The output is in files {\em dipnrratesm.dat} and {\em dipnrratesv.dat},
which contains non-radiative rates normalized to the radiative rates
in a corresponding infinite space.

The output files {\em *m.dat} and {\em *v.dat} differ 
only in that the rates are normalized with respect to those in the local medium
at the dye position and in vacuum respectively.
Each output file has data organized in four columns: 
normalized radial position (in the units of the bead radius) 
and the corresponding quantities for the 
respective tangential, radial, and averaged dipole orientations 
(from left to right). [The fourth column is 
not a simple average of the second and third columns but 
is calculated according to Eq. (123) of Ref. \cite{AMap} 
as $(\Gamma_\perp + 2 \Gamma_\parallel)/3$)]. To modify the default versions 
of the two codes to the present case of a homogeneous sphere, 
the following steps have to be performed: 
 
\begin{itemize}

\item
The layers of a coated sphere with N shells are counted from 
one for the sphere core up to LCS=N+1 for the outermost shell 
and N+2 for the host medium.
See also Fig. 1 of \cite{AMap}. A homogeneous sphere has no shell and thus N=0.
Consequently, the value of the integer parameter LCS, which controls
the number of coatings, has to be set to LCS=1. Next 
the integer parameter ILCS has to be set to ILCS=1. The latter ensures that 
the dielectric constant specified by the complex parameter CCEPS
becomes the dielectric constant of the sphere (core).

\item 
The value of the complex parameter CCEPS has to become
one of the two values of the complex dielectric constant.
At the same time the value of the integer parameter
NMAT=0 has to be adopted. The latter choice corresponds to a 
dispersionless medium and should also be imposed if the data
specified by CCEPS is given preference. 
Otherwise, for NMAT $\geq 1$,
the resulting value of the dielectric constant, which enters
the Mie coefficients as the complex variable ZEPS1, is 
overwritten by either the Drude data or the real material data.

\item The integer parameter NABS, which specifies the total 
number of absorbing layers,
 and IABS, which carries the number of the absorbing layer, 
have to be both set to one.

\item The value of the real parameter NX specifies in the 
units of the sphere radius the length of the radial 
interval on which the decay rates are calculated. The value
of NX is used to control the ``do 200 il=1,nx" loop.

\item
The angular momentum number $l$ of the induced multipoles involved 
in the calculation is controlled through the value of the variable LMX. 
The data setting LMX/1/
then forces the code to take into account only the dipolar
contribution. The arrays of the codes are dimensioned with 
the integer parameter LMAX, where at present LMAX=60.
The variable LMX can take any value between 1 and LMAX.

\end{itemize}

\section{Supporting Information Available}
A MS Windows executable \cite{exe} is provided which calculates 
all the approximate rates shown here.


\newpage


\begin{center}
{\large\bf Figure captions}
\end{center}

\vspace*{0.4cm}

\noindent {\bf Figure 1 -} 
Normalized fluorescence decay rates versus the distance $z$ to the center of a AgNP 
of radius $a=5$ nm for the emission wavelength $\lambda=612$ nm.
The exact rates shown by a solid black line here and in other figures of the paper
comprise the contribution of the induced multipoles 
with $l$ up to $l_{max}=50$. The exact rates with an imposed cut-off at
$l=1$ are shown by a dash (red) line. The approximate dipole-dipole
interaction results, irrespective if obtained 
with $\alpha_C$ (dash-dot line; blue),
$\alpha_{Mie;as}$ (dot line line; green), or
 $\alpha_{Mie}$ (dash-dot-dot line; cyan), merge to a single line with that for 
the exact rates with a cut-off imposed at $l=1$.

\vspace*{0.4cm}

\begin{figure}[tbp]
\begin{center}
\epsfig{file=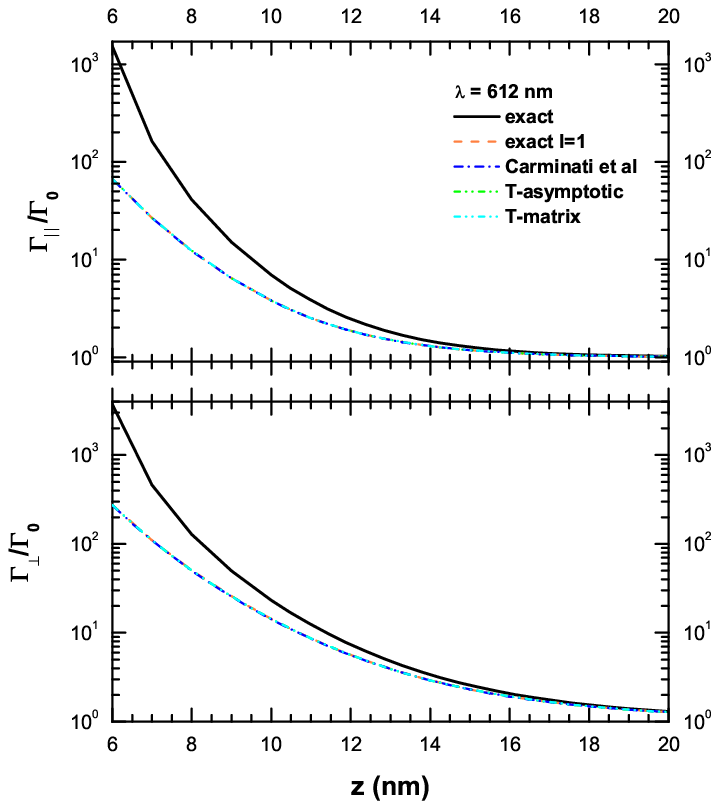,width=14cm,clip=0,angle=0}
\end{center}
\caption{}
\label{qm612}
\end{figure}

\begin{figure}[tbp]
\begin{center}
\epsfig{file=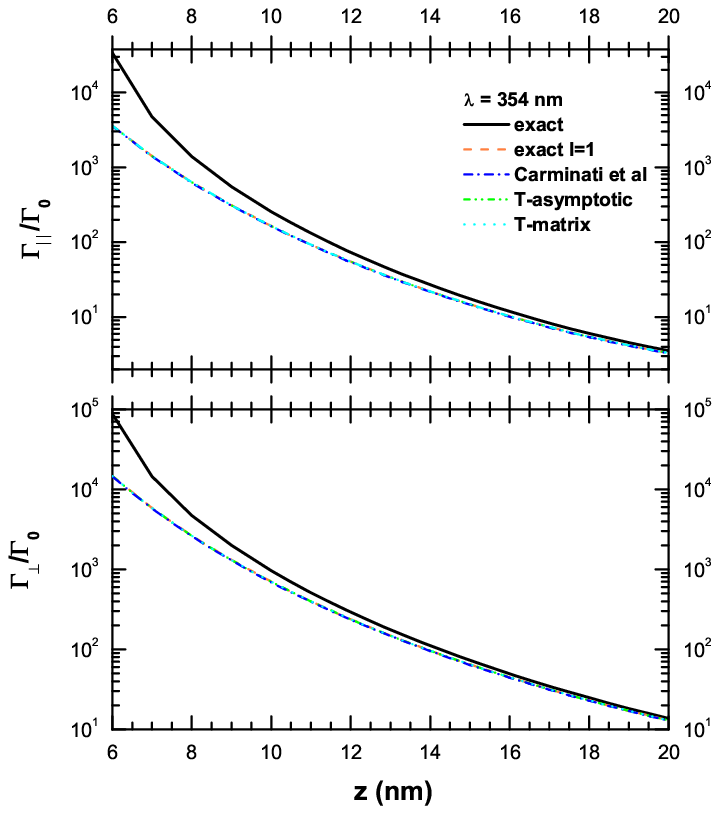,width=14cm,clip=0,angle=0}
\end{center}
\caption{}
\label{qm354}
\end{figure}

\noindent {\bf Figure 2 -} 
The same as in Fig. \ref{qm612} but 
for the emission wavelength $\lambda=354$ nm.

\vspace*{0.4cm}

\noindent {\bf Figure 3 -} 
Normalized radiative rates versus the distance $z$ to the center of a AgNP 
of radius $a=5$ nm for the emission wavelength $\lambda=612$ nm.
The approximate dipole-dipole
interaction results, irrespective if obtained 
with $\alpha_R$ (short dash line; mangenta), $\alpha_C$ (dash-dot line; blue),
$\alpha_{Mie;as}$ (dot line line; green), or
 $\alpha_{Mie}$ (dash-dot-dot line; cyan), merge to a single line with 
that for the exact radiative rates. There is essentially
no difference if, in the latter case, the cut-off is imposed 
at either $l=1$ (dash line; red) or $l=50$ (solid line; black).

\vspace*{0.4cm}

\begin{figure}[tbp]
\begin{center}
\epsfig{file=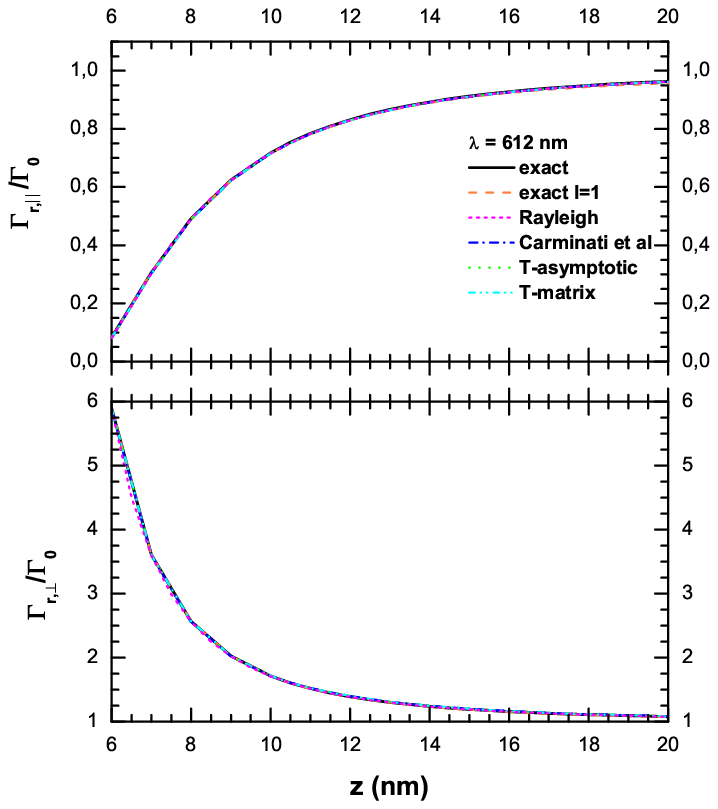,width=14cm,clip=0,angle=0}
\end{center}
\caption{}
\label{rrad612}
\end{figure}

\begin{figure}[tbp]
\begin{center}
\epsfig{file=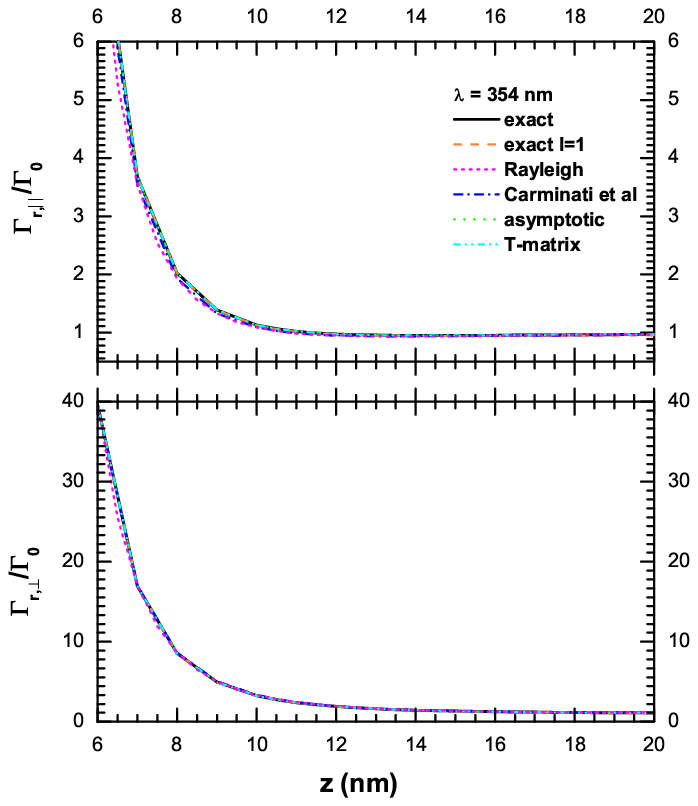,width=14cm,clip=0,angle=0}
\end{center}
\caption{}
\label{rrad354}
\end{figure}

\begin{figure}[tbp]
\begin{center}
\epsfig{file=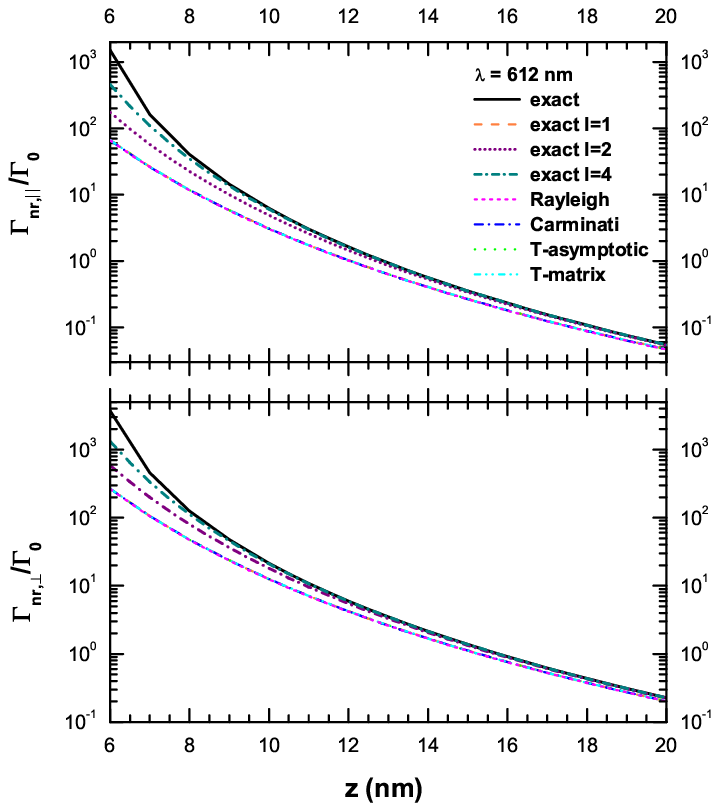,width=14cm,clip=0,angle=0}
\end{center}
\caption{}
\label{nrad612}
\end{figure}

\begin{figure}[tbp]
\begin{center}
\epsfig{file=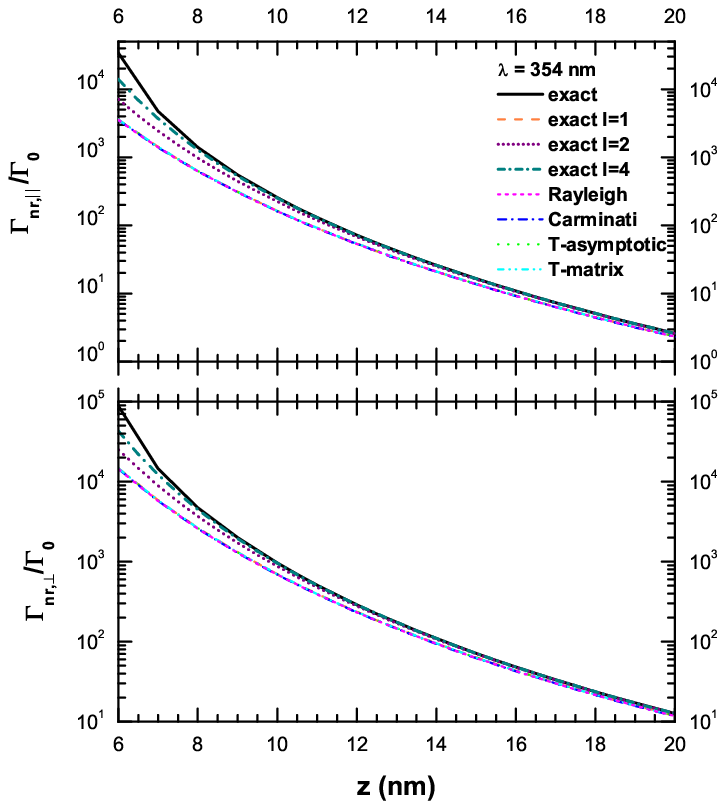,width=14cm,clip=0,angle=0}
\end{center}
\caption{}
\label{nrad354}
\end{figure}

\begin{figure}[tbp]
\begin{center}
\epsfig{file=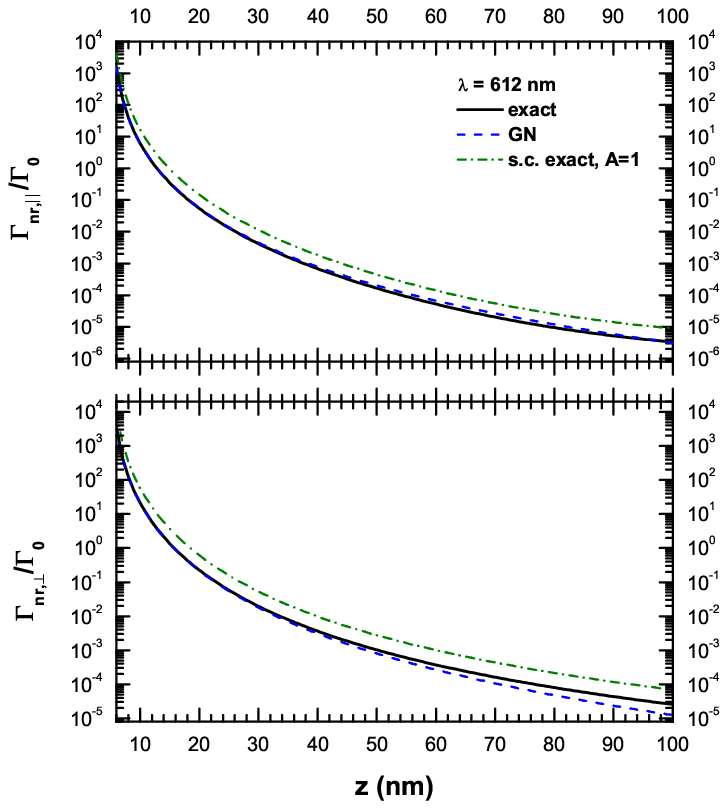,width=14cm,clip=0,angle=0}
\end{center}
\caption{}
\label{gnnrex612}
\end{figure}

\begin{figure}[tbp]
\begin{center}
\epsfig{file=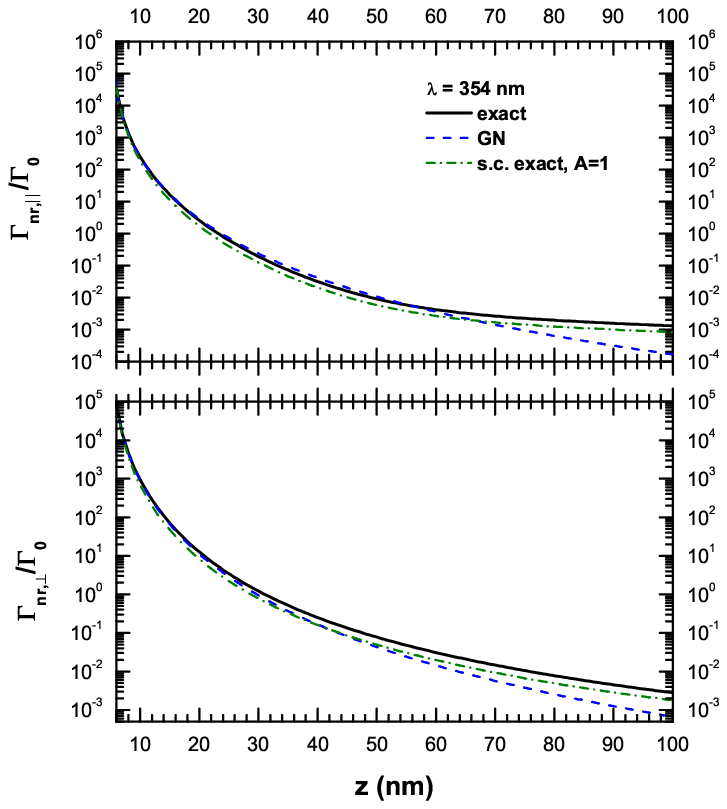,width=14cm,clip=0,angle=0}
\end{center}
\caption{}
\label{gnnrex354}
\end{figure}

\noindent {\bf Figure 4 -}
The same as in Fig. \ref{rrad612} but 
for the emission wavelength $\lambda=354$ nm.

\vspace*{0.4cm}

\noindent {\bf Figure 5 -} 
Normalized non-radiative rates versus the distance $z$ to the center of a 
AgNP of radius $a=5$ nm for the emission wavelength $\lambda=612$ nm.
The approximate dipole-dipole
interaction results, irrespective if obtained 
with $\alpha_R$ (short dash line; mangenta), $\alpha_C$ (dash-dot line; blue),
$\alpha_{Mie;as}$ (dot line; green), or
 $\alpha_{Mie}$ (dash-dot-dot line; cyan), merge to a single line with that for 
the exact radiative rates with a cut-off imposed at $l=1$ (dash line; red).
The figure demonstrates a clear difference between
the approximate non-radiative rates and exact non-radiative rates with a cut-off 
imposed at $l=2$ (short dot line; purple), $l=4$ (short dash-dot line; dark cyan), 
and $l=50$ (solid line; black), respectively.

\vspace*{0.4cm}

\noindent {\bf Figure 6 -}
Normalized non-radiative rates versus the distance $z$ to the center of a 
AgNP of radius $a=5$ nm for the emission wavelength $\lambda=354$ nm. 
The exact rates with an imposed cut-off at
$l=1$ are shown by a dash (red) line.

\vspace*{0.4cm}

\noindent {\bf Figure 7 -} 
Comparison of the normalized Gersten and Nitzan non-radiative rates 
against the exact  rates for a AgNP of radius $a=5$ nm for 
the emission wavelength $\lambda=612$ nm.
The exact results with a size corrected dielectric function
of silver for $A=1$ are shown by dash-dot line (olive).
All the rates were calculated with a cut-off $l_{max}=50$.

\vspace*{0.4cm}

\noindent {\bf Figure 8 -}
Comparison of the normalized Gersten and Nitzan non-radiative rates 
against the exact rates for a AgNP of radius $a=5$ nm
for the emission wavelength $\lambda=354$ nm. 
The exact results with a size corrected dielectric function
of silver for $A=1$ are shown by dash-dot line (olive).
All the rates were calculated with a cut-off $l_{max}=50$.

\newpage

\newpage

\newpage

\newpage

\newpage

\newpage

\newpage

\end{document}